\documentstyle[amssymb,preprint,aps]{revtex}
\begin{document}
\draft
\tightenlines 
\author{V.N. Kotov\cite{add}, J. Oitmaa\cite{j}, and Zheng Weihong\cite{z}}
\address{School of Physics, The University of New South Wales \\
Sydney 2052, Australia}
\title{Excitation spectrum and ground state properties of the 
$S=\case 1/2$ Heisenberg ladder with staggered dimerization}  
\date{
\today%
}
\maketitle

\begin{abstract}
We study  the
 excitation spectrum and ground state properties 
of the two-leg $S=\case 1/2$ quantum spin ladder with
 staggered dimerization.  Two massive phases, separated by a critical 
 line are found, as predicted by previous analysis, based on the
 non-linear sigma model (NLSM). We have used  dimer
series expansions, exact diagonalization of small clusters
 and diagrammatic 
analysis of an effective interacting Bose gas
 Hamiltonian, obtained by using the bond operator representation for
 spins. We compute the excitation spectrum in the phase, 
dominated by strong rungs in the parameter regimes far
 away and close to the point
 of instability.
The  exact location of the phase boundary  is determined.   
\end{abstract}

\pacs{75.10.Jm, 75.30.Ds, 75.40.Gb, 75.30.Kz}

\baselineskip 0.6cm

\section{Introduction}

There is a great deal of 
current interest in quasi one-dimensional quantum spin models, such as
 spin ladders,  spin chains with dimerization and (or)
 frustration, and various generalizations of the above. The 
studies of these models have been mostly triggered by the experimental
 discoveries of  several spin-Peierls and spin-ladder compounds \cite{review}.
 The theoretical efforts have been focused mainly on the nature of the
ground state, the  excitation spectrum, as well as thermodynamic
properties of these models. The system of coupled quantum $S=\case 1/2$ chains 
 (spin-ladder) was found to behave  differently
 depending on whether the number of chains is even or odd \cite{review}.     
 For an even number,   the  excitation spectrum 
 is generically gaped and
short-range correlations dominate in the ground state. 
 The single $S=\case 1/2$ quantum spin chain with dimerization exhibits quite similar
 properties, since it is known that dimerization produces an energy
gap and destroys the quasi long-range order of the integrable
 uniform chain. 

In a recent paper,  Ref.\cite{snake}, M.A. Martin-Delgado, R. Shankar, and
G. Sierra 
  have proposed a model of coupled spin chains with staggered
 dimerization  which  possesses a rich phase diagram, quite different
 from that of the simple spin ladder (no dimerization). Consider the
simplest version of the model, containing two coupled $S=\case 1/2$
 Heisenberg chains  (Fig. 1.):

\begin{equation}
 H = J_{\perp} \sum_{i} {\bf S}_{1,i} \cdot {\bf S}_{2,i} + 
  \sum_{i} \sum_{\alpha=1,2}
 J[1+ \delta (-1)^{i+\alpha}]{\bf S}_{\alpha,i}\cdot {\bf S}_{\alpha,i+1}
\end{equation}
Here $\alpha=1,2$ is labeling the two chains (legs of the ladder).
 The coupling 
 $J_{\perp}\geq0$ is the inter-chain (rung) interaction and $J(1\pm
\delta)\geq0$ are the nearest neighbor interactions along the chains
(all couplings are antiferromagnetic). 
Notice that the dimerization $\delta$
 is staggered in both directions (along and perpendicular to the  
chains), i.e. the dimerization in each chain is in antiphase with its
neighbor chain.
The model Eq.(1) was analyzed in Ref.\cite{snake} by mapping onto
a non-linear sigma model (NLSM).
  In the NLSM
approach the value of the topological angle $\theta$ determines
 the nature of the possible phases of the model \cite{affleck}.
 For $\theta=0$ the model is massive
while it is massless for $\theta=(\mbox{odd multiple of} \ \pi)$.
 For the model at hand two massive (gaped) phases 
 were found in the parameter plane $(\delta, J_{\perp}/J)$ with
 a massless (critical) line separating them, determined by the
equation:

\begin{equation}
 \theta =\frac{4 \pi \delta}{2+ J_{\perp}/2J}= \pi \ \ \Leftrightarrow
\ \
 4\delta = 2+ J_{\perp}/J \ \ (\mbox{critical line}).
\end{equation}
A symmetric branch with $\delta \rightarrow -\delta 
\ (\theta=3\pi)$ also exists
 but we will assume $\delta\geq0$ from now on.
 The two massive phases separated by this line are basically
phases where the dominant correlations are along the rungs or
 the stronger bonds along the legs. The existence of a critical
line is quite unusual  since individually
 both interactions, $J_{\perp}$ and $\delta$ are gap producing.

It is clear from Fig. 1. that there are two points
 $(\delta,J_{\perp}/J)= (0,0)$ and $(1,2)$ where the system is a single
uniform chain and hence integrable and massless. The NLSM approach
 Eq.(2) does not correctly reproduce the point $(0,0)$ which is
quite natural since it is expected to work in the vicinity of the
 point  $(1,2)$ only. It has been conjectured in Ref.\cite{snake}
 essentially from continuity, that a critical line connects the
 two integrable points. 

Let us mention that  upon introducing an additional spin alternation 
\cite{kawakami} 
 and (or) considering ladders of spin $S=1$ or higher
\cite{snake} 
the phase diagram of the system will be even more complex.
 The basic origin for this complexity however is still the 
 staggered dimerization in the rung direction. 
Thus, form now on  we will consider only the ``minimal" model Eq.(1).

 The purpose of the present work  is to analyze the
 excitation spectrum and the phase diagram  
of the model  Eq.(1) at $T=0$
 by a combination of numerical and analytical
techniques.  In Sec. II we present exact diagonalization results
 for the energy gap and the spin-spin correlations, confirming the
 existence of two gaped phases separated by a gapless line. 
 In Sec. III  we develop two kinds of strong-coupling approaches 
 in the phase dominated by strong interchain (rung) correlations:
(a) dimer series expansions, and (b) diagrammatic treatment of 
 the effective Hamiltonian, written in terms of bond triplet
 operators. The spectrum  near the critical line is calculated
 and the relevant interactions, responsible for the closing of
 the gap are identified.  
Sec. IV contains our conclusions and discussion of 
future prospects.

\section{Exact diagonalizations}
 First we present results from finite lattice diagonalizations 
of the Hamiltonian Eq.(1), for systems of up to $N=24$ spins.              
Periodic boundary conditions are used throughout.
 For reasons of symmetry the number of spins must be a multiple
of $4$, and most results are obtained from extrapolation
 of $12,14,16,20,24$ spin systems. We find empirically that
 the ground state is a singlet $S_{{\rm tot}}=0$ whereas the first
excited state is a triplet $S_{{\rm tot}}=1$. The computed singlet-triplet
gaps are extrapolated to the thermodynamic limit $N \rightarrow \infty$
via the ansatz:

\begin{equation}
\Delta E_{N} = \Delta E_{\infty} + \frac{A}{N} + \frac{B}{N\ln N}
\end{equation} 
where $A$ and $B$ are constants. 
The last term allows for the logarithmic corrections arising
 through conformal invariance in spin chains and effectively
 accounts for the residual curvature observed in direct ratio plots.

 In Fig. 2. we show the variation of energy gaps versus the
 interchain coupling $J_{\perp}/J$ for various $N$, for a fixed
 value of the dimerization $\delta$ (in this case $\delta=\case 1/2$).
It is clear qualitatively  that the variation in
 $\Delta E_{N}$ is small for both small and large values of 
 $J_{\perp}/J$ whereas the variation is large near  $J_{\perp}/J
\approx 1.2$ where the individual curves have a minimum. The dashed
lines, based on the extrapolated formula (3) clearly show the gap
vanishing at a single point on the $J_{\perp}/J$ axis. Similar results
are obtained for other values of $\delta$ although for small  
 $\delta$ the convergence is poorer. This is consistent with
 the increasing quantum fluctuations and larger finite size        
corrections expected near the uniform limit. This analysis leads to
 Fig. 3. which shows the estimated critical line in the 
$(\delta,J_{\perp}/J)$ plane. Comparing with Eq.(2) one can
see that the NLSM result can not be trusted numerically, but
 the existence of a gapless line is correctly predicted.
 
In Fig. 4. we show various spin correlations versus  $J_{\perp}/J$ 
 for  $\delta=\case 1/2$. As intuitively expected the phase
 with $J_{\perp}/J > 1.22$  is characterized by strong 
 inter-chain (rung) and weak intra-chain correlations.
 When $J_{\perp}/J$ decreases and ultimately crosses 
the critical point, the behavior of the correlation functions
is reversed, signaling a transition into a phase, dominated by
strong intra-chain correlations on the stronger bonds.
 We find that the spin correlations as well as the excitation
 gap change continuously through the quantum transition point.        

\section{Strong coupling expansions}
 In this section we present analysis based on strong-coupling
expansions around the limit $J_{\perp} \gg J$. 
Two approaches are used:  the dimer series expansion and
the bond operator  mapping onto an effective interacting Bose gas.
 The two approaches are similar in spirit and produce quite similar
 result even though the technical details  are different.

\subsection{Dimer series expansions} 
We start with results obtained by the linked-cluster 
dimer series expansion method. The linked-cluster expansion method
has been previously reviewed in several
articles\cite{he90,singh,gelmk}, and will not be repeated here.
The basic idea is to divide the Hamiltonian into two parts:
\begin{equation}
H = H_0 + V
\end{equation}
where
\begin{eqnarray}
H_{0} & =& J_{\perp} \sum_{i} {\bf S}_{1,i}\cdot {\bf S}_{2,i}, \nonumber \\ && \\
V &=& J \sum_{i} \sum_{\alpha=1,2}
 [1+ \delta (-1)^{i+\alpha}]{\bf S}_{\alpha,i}\cdot {\bf S}_{\alpha,i+1} \nonumber
\end{eqnarray}
By fixing values of $\delta$, we can construct an expansion 
in $J/J_{\perp}$ by taking  $H_0$ as 
unperturbed Hamiltonian and $V$ as a perturbation.  The zeroth order $V=J=0$
 approximation corresponds to isolated dimers (singlets) on
 the rungs. Each singlet can be excited into a triplet
 state and thus the excitation gap is $J_{\perp}$. Finite $J$
 introduces interactions between the rungs which modifies
 substantially the spectrum. In order to consider the physically
 interesting coupling region $J_{\perp}/J \sim 1$ the perturbation
series has to be developed to high order and then extrapolated to
the relevant point. We have carried out an expansion 
up  to order $(J/J_{\perp})^{23}$ for the ground state energy,  to
order $(J/J_{\perp})^{13}$ for antiferromagnetic susceptibility, and
to order $(J/J_{\perp})^{11}$ for the triplet excitation spectra for several
values of $\delta$.
There are only 12 graphs that contribute to the ground state energy and
dispersion, and 14 graphs that contribute to the 
antiferromagnetic susceptibility. 
The series coefficients are not presented here,
but are available upon request.
Then we have used integrated 
 differential approximants and Pad\'e approximants\cite{tony} 
to extrapolate the series. It is sometimes useful
to compare the present results with the spectrum of a 
 spin ladder without dimerization 
$(\delta=0)$, which can be found in Ref. \cite{series}
(see Fig. 4.). 

In Fig. 5. we present the triplet excitation spectrum $\omega(k)$ for
$\delta=\case 1/2$, obtained by the dimer series
expansions. The triplet gap  $\Delta=\omega(\pi)$  decreases
 with the decrease of $J_{\perp}/J$, vanishing (within error bar)
 at the  critical value $J_{\perp}/J=1.23$.  
 By comparing the spectra of Fig. 5. with the corresponding
 spectra for $\delta=0$ (see Fig. 4. of Ref.\cite{series} for the value
 $J_{\perp}/J=2$) one can see that  the renormalization
 due to finite  $\delta$ is the strongest at $k=\pi$ while
 in the vicinity of the point $k=0$ the energy is almost 
unrenormalized.  This is due to the fact that the first
three leading orders in $J/J_{\perp}$ do not change $\omega(k=0)$
 (see also Sec. III.B). 

 A more accurate determination of the critical line can be
 achieved by the Dlog Pad\'e approximants to the 
antiferromagnetic susceptibility series,
 and the results are presented in Fig. 3. 
 The  phase boundary
 obtained within the dimer series approach is in excellent
 agreement with  the exact diagonalization results
 in the region $J_{\perp}/J > 1$. For small values
 of $J_{\perp}$ the convergence of the dimer series becomes  poorer 
  (larger error bars), as the critical point occurs at much larger $J/J_{\perp}$.   
 In Fig. 6. several critical spectra, for parameters on the massless
 line, are plotted. Notice that for $\delta=1, J_{\perp}=2J$ when
 Eq.(1) reduces to a ``snake", i.e. one-dimensional Heisenberg chain,
 the dimer series reproduces correctly the spinon dispersion,
 which is known from the exact Bethe ansatz solution of the problem  
\cite{spinon}. The latter is given by the formula (for a chain
 with exchange $J$):  
 $\omega(k)/J=(\pi/2)\sin(\frac{\pi-k}{2})$.  Indeed, the
 lower curve in Fig. 6. practically coincides with this formula.
 
Using the long series for the ground state energy, 
we were able to obtain the most accurate
estimates of the ground state energy, as it was done before for the normal 
two-chain ladder\cite{ladder2}.
For example for $\delta=0.5$ and $J/J_{\perp}=0.8$ which is 
near the critical line,
the ground-state energy per site is estimated to be
\begin{equation}
E_0/NJ_{\perp} = -0.52655(2)~.
\end{equation}

Beside the above inter-chain 
dimer expansions about the limit $J_{\perp} \gg J$,
we can also construct another dimer expansion, the intra-chain dimer expansion,
 by taking the stronger intra-chain bonds
(the bonds couping $J(1+\delta)$) as unperturbed Hamiltonian, and the rest of 
the bonds as perturbation, and carry out dimer series expansions in $(1-\delta)/(1+\delta)$
for fixed values of $J_{\perp}/J(1-\delta)$. The series have been computed  
up to order $[(1-\delta)/(1+\delta)]^{11}$ for the
ground state energy, to order $[(1-\delta)/(1+\delta)]^{6}$
for the antiferromagnetic susceptibility, and to order $[(1-\delta)/(1+\delta)]^{7}$ for 
the triplet excitation spectra for several values of $J_{\perp}/J(1-\delta)$.
With this expansion, one can study the excitation spectrum and 
the ground state properties for the 
parameters located within the intra-chain dimer phase.
The gapless critical line derived from this expansion agrees well with that by
inter-chain dimer expansions, but with much less accuracy. 
Here we will not present any detailed results from this dimer expansion,
however the series are available upon request.
  
\subsection{Bond operator representation}

  In this section we describe a mapping of the model onto an
interacting Bose gas, which can be achieved by using the
bond operator representation for spins \cite{subir}. 
 Denote by $|0\rangle_{i}$ the singlet state formed by
 two spins on rung $i$. Then, in the strong-coupling limit
$J_{\perp}/J \gg 1$  the excitations are well described by triplets, created
 by the (bond) operators $t_{i,\alpha}^{\dag}$, 
 $|\mbox{triplet}\rangle_{i} = t_{i,\alpha}^{\dag}|0\rangle_{i}$,
where $\alpha = x,y,z$ are the three components of the triplet 
and the usual bosonic commutation relations hold
 \cite{subir}.     
 The Hamiltonian Eq.(1) expressed in terms of the triplet 
operators is \cite{remark}:    

\begin{eqnarray}
\label{H1}
H&=&\sum_{i,\alpha,\beta}\left\{J_{\perp}t_{\alpha i}^{\dag}t_{\alpha
i}+
{{J}\over{2}}\left(t_{\alpha i}^{\dag}t_{\alpha i+1}+
t_{\alpha i}^{\dag}t_{\alpha i+1}^{\dag}+\mbox{h.c.}\right) \right.
\nonumber \\
& & \left. + {{J}\over{2}}\left(t_{\alpha i}^{\dag}
t_{\beta i+1}^{\dag}t_{\beta i}t_{\alpha i+1}
-t_{\alpha i}^{\dag}t_{\alpha i+1}^{\dag}t_{\beta i}t_{\beta
i+1}\right)
\right\} + H_{\delta} + H_{U},
\end{eqnarray}
 where we have separated the part due to dimerization:
\begin{equation}
H_{\delta} =\frac{ \delta}{2} \sum_{i,\alpha,\beta,\gamma}
(-1)^{i} \left \{\left [i \epsilon_{\alpha\beta\gamma}
t_{\alpha i+1}^{\dag} t_{\beta i}^{\dag}  t_{\gamma i} +
\mbox{h.c.} \right ] + [i \leftrightarrow i+1] \right \}
\end{equation}
and $H_{U}$ will be defined below.
For $\delta=0$ the Hamiltonian (7) coincides with that of
 the simple two-leg ladder \cite{Gopalan,us2}. 
 In order for the triplet operators to represent only the physical
 spin triplet states, they have to satisfy the on-site hard-core
 constraint: $t_{\alpha i}^{\dag}  t_{\beta i}^{\dag}=0$. 
 This restriction on the Hilbert space can be taken into account by introducing
 an infinite on-site repulsion between the triplets, as discussed
 in our previous work \cite{us2,us1}: 

\begin{equation}
H_{U} = U \sum_{i,\alpha \beta}
t_{\alpha i}^{\dagger}t_{\beta i}^{\dagger}t_{\beta i}t_{\alpha i},
 \ \ U \rightarrow \infty.
\end{equation} 
 Our goal is to develop a diagrammatic treatment of 
the interactions  in the Hamiltonian (7) in order 
 to understand which terms  are most relevant to the renormalization
 of the spectrum. The quadratic part (first line in Eq.(7)) is
diagonalized in momentum space by a Bogoliubov transformation 
 $t_{\alpha k} \rightarrow u_{k}t_{\alpha k} +
 v_{k}t_{\alpha -k}^{\dag}$, which leads to the spectrum
$\omega^{2}_{k} = A_{k}^{2} - B_{k}^{2}$, with 
$A_{k}= J_{\perp} + J \cos(k), \ B_{k} =  J \cos(k)$.  
 The Bogoliubov coefficients are defined as:
 $u_{k}^{2}, v_{k}^{2} = \pm 1/2 + A_{k}/2\omega_{k}$.

The  spectrum of the Hamiltonian Eq.(7) without  the dimerization
 term $H_{\delta}$ was investigated diagrammatically 
in Refs.\cite{us2,us1}. Here we only summarize the results. 
  The on-site repulsion $H_{U}$  gives the dominant
 contribution to the spectrum renormalization, while the two-particle
inter-site interaction with strength $J/2$ (first term in the second line of
 Eq.(7)) is a relatively minor effect \cite{remark1}.
 Even though the on-site interaction is infinite, the
 scattering amplitude $\Gamma_{\alpha \beta, \gamma \delta} (k, \omega) 
 = \Gamma(k, \omega) (\delta_{\alpha \gamma} \delta_{\beta \delta}
+ \delta_{\alpha \delta} \delta_{\beta \gamma})$
is finite (as physically expected), 
 which can be seen by resumming the ladder
 series, shown in Fig. 7(a) and setting $U \rightarrow \infty$.
 The scaterring amplitude and the corresponding self-energy  
 (Fig. 7(b)) are  \cite{us2,us1}:

\begin{equation}
[\Gamma(k,\omega)]^{-1} = \sum_{q}\frac{1}{-\omega + \omega_{q}+
\omega_{k-q}} \ , \ \Sigma_{U}(k,\omega) = 4 \sum_{q} v_{q}^{2} \Gamma(k
+q,\omega - \omega_{k}).
\end{equation} 
Here, and in all future equations   we set, for convenience, the number of rungs to unity. 
 As emphasized in Ref.\cite{us2,us1},  Eq.(10) represents 
 the dominant contribution from $H_{U}$, provided the
 density of triplets $N_{t}= \langle 
 t_{\alpha, i}^{\dagger} t_{\alpha, i} \rangle =
 3\sum_{q}v_{q}^{2}$ is small.
 Thus Eq.(10) should be viewed as the first term in an 
expansion in powers
 of  $N_{t}$. For the simple ladder ($\delta=0$) we find
 (after solving the corresponding Dyson equation, see below)
 $N_{t}\approx 0.1$ for $J_{\perp}=2J$, and
 $N_{t}\approx 0.25$ for  $J_{\perp}=J$.
 Therefore the dilute gas approximation is expected to work
 quite well 
 even for  $J_{\perp}/J \sim 1$.
 Naturally when $J_{\perp}/J \rightarrow 0$ the above picture, based
 on strong rungs, breaks down. Instead, the excitations in this regime
 should be viewed as weakly bound spinons, thus making the local
triplets an inadequate staring point.  
 We have estimated that the bond operator formalism              
 describes well the excitation spectrum for $J_{\perp}/J > 1$
 \cite{us2,us3}.

Next we turn to the three-particle term $H_{\delta}$
 which represents the dimerization part of the Hamiltonian.
 To lowest (one-loop) order there are contributions to the
 normal and anomalous self-energies, drawn in Fig. 7(c) and
  Fig. 7(d), respectively. The corresponding formulas are:

\begin{equation}
\Sigma_{\delta}^{N}(k,\omega) = 4 (\delta J)^{2}
\sum_{q} \frac{[C(q+\pi,k-q,k)]^{2}}{\omega - \omega_{q+\pi} - 
\omega_{k-q}} +
\left\{ \begin{array}{c} u \leftrightarrow v \\
\omega \rightarrow -\omega \end{array} \right\},
\end{equation}

\begin{equation}
\Sigma_{\delta}^{A}(k,\omega) = 4 (\delta J)^{2}
\sum_{q} \frac{C(q+\pi,k-q,k)D(q+\pi,k-q,k)}{\omega - \omega_{q+\pi} -
\omega_{k-q}} +
\left\{ \begin{array}{c} u \leftrightarrow v \\
\omega \rightarrow -\omega \end{array} \right\},
\end{equation}
 where the following definitions are used 

\begin{equation}
C(k,p,q) = \Phi(k,p)u_{k}u_{p} + \Phi(-q,p)u_{k}v_{p}
 +  \Phi(k,-q)v_{k}u_{p},
\end{equation}
\begin{equation}
D(k,p,q) = - C(k,p,q) \{u \leftrightarrow v\}, 
\Phi(k,p) = \frac{1}{2}(\sin k - \sin p).
\end{equation}
The function $\Phi(k,p)$ originates from the Fourier transform
 of the three-particle vertex Eq.(8). Equations (11) and (12) can be
 derived by evaluating all possible internal loops in the diagrams
 in Figs. 7(c),(d), where, as usual, lines with a single arrow stand for the
 normal Green's function $G^{N}(k,t)= -i\langle T t_{\alpha,k}(t)
t^{\dag}_{\alpha,k}(0) \rangle$  and lines with double arrows
 represent the anomalous Green's functions 
  $G^{A}(k,t) = -i\langle T t^{\dag}_{\alpha,-k}(t) t^{\dag}_{\alpha,k}(0)
\rangle$.
 Finally, the coupled Dyson's equations for the normal
 and anomalous Green's functions have to be solved leading to the
 result (more details can be found in \cite{us3}):

\begin{equation}
G^{N}(k,\omega) = \frac{\omega + A_{k} + \Sigma^{N}(k,-\omega)}
{[\omega + A_{k} + \Sigma^{N}(k,-\omega)]
[\omega -  A_{k} - \Sigma^{N}(k,\omega)] + 
[B_{k} + \Sigma^{A}(k,\omega)]^{2}}, 
\end{equation}
where the total normal and anomalous self-energies are defined as
\begin{equation} 
\Sigma^{N}(k,\omega) = \Sigma_{U}(k,\omega) + \Sigma_{\delta}^{N}(k,\omega),
\ \  \Sigma^{A}(k,\omega) = \Sigma_{\delta}^{A}(k,\omega).
\end{equation}

In order to  get a feeling for the effect of $H_{\delta}$ on the
 spectrum, it is instructive to examine the second order 
perturbation theory result. It can be obtained by
 noticing that to leading order $\omega_{k} \approx J_{\perp},
 u_{k} \approx 1, v_{k} \approx -(J/2J_{\perp}) \cos k$. Thus
 in evaluating the leading order correction arising from
 the dimerization term one can set $v_{k}=0$ which means that
 only the normal self-energy contributes:  
\begin{equation}
\delta \omega_{k} = \Sigma_{\delta}^{N}(k, \omega_{k}) =  
4 (\delta J)^{2}\sum_{q} \frac{\Phi^{2}(q +\pi, k-q)}{-J_{\perp}} =  
 \frac{\delta^{2} J^{2}}{J_{\perp}}(\cos k -1).
\end{equation}
One can see that to lowest order the renormalization of the
 spectrum for $k=\pi$ is the strongest while $k=0$ is not
 renormalized. For a fixed value of $\delta$, the gap
 $\omega(\pi)$ decreases with decreasing $J_{\perp}/J$ and
 vanishes at a critical ratio, signaling an instability
 of the rung dimer phase. 

 We have found the spectrum  numerically by solving for the poles
 of the exact Green's function Eq.(15), thus resumming the
perturbation series to all orders. The results 
 for $\delta=\case 1/2$ are presented in
 Fig. 5. The diagrammatic results
are  in  good agreement with the dimer series expansions, however
with  
 decreasing $J_{\perp}/J$ the disagreement  increases.
 We would like to point out that even though  the two methods have the
 same starting point (strong rungs) and typically agree quite
 well in the region of strong coupling, for smaller coupling the
 dimer series expansion performs better. The reason is that
 the series expansion takes 
 into account all terms  to a particular (finite)
 order, while the diagrammatic approach sums up  only  the most
 dominant subclass of diagrams. The small parameter
 $N_{t}$ controlling the diagrammatic expansion grows with 
 decreasing $J_{\perp}/J$ and at $J_{\perp}=1.23J$ (the critical
 point, as estimated by the dimer series and exact diagonalization)
 has the value $N_{t} = 0.3$. 
  This relatively large value of the triplet density is related
 to the considerably decreased magnitude of 
the spin-spin correlation function on the
rungs\cite{remark2} (see the (1,2) correlations in Fig. 4.). 
 Therefore the agreement between 
 the diagrammatic analysis and the dimer series is, in fact, even better
 than expected.   
  
\section{Summary and discussion}
 
 In conclusion, we have investigated the excitation spectrum and
 the ground state phase diagram of the quantum spin ladder
 with staggered dimerization and have found that a transition
 takes place form a phase dominated by strong rung correlations
 into a phase, characterized by intrachain dimers. The two phases
 are separated by a gapless line. The location of the boundary, 
 the triplet excitations, as well as the spin-spin 
correlations in the rung dimer phase were calculated by 
 dimer series expansions, diagrammatic analysis of 
 the effective hard-core Bose gas of rung triplets, and exact
diagonalizations. 

We have identified the three-particle
 scattering $H_{\delta}$, in terms of the rung triplets, to
 be responsible for the instability of the rung dimer phase  
 (closing of the gap). It is interesting to note
 that this is indeed the term which, in the continuous
 field theory formulation, contributes to the topological
 angle $\theta$ and leads to the transition. 
 Indeed, from the representation of the spin operator 
 on a given rung $i$ 
 in terms of the triplets \cite{remark} one can easily see
 that ${\bf S}_{1} - {\bf S}_{2} \sim {\bf t}_{i} + 
 {\bf t}^{\dagger}_{i} \sim \mbox{\boldmath $\phi$}_{i}$,
  ${\bf S}_{1} +  {\bf S}_{2} \sim i  {\bf t}^{\dagger}_{i}
\times {\bf t}_{i}  \sim {\bf l}_{i}$. These are precisely the
 two vector fields used in the derivation of the non-linear sigma model
 \cite{affleck}. Therefore the terms of the form
 $\mbox{\boldmath $\phi$}_{i} \cdot
{\bf l}_{i+1}$  which appear in our lattice 
 Hamiltonian $H_{\delta}$ would be precisely the ones  
 which give a non-trivial $\theta$ after the continuum limit
 is taken.

  There are  several additional aspects of the problem that can be addressed 
 within the strong-coupling formalism described in
 this paper. As suggested in Ref.\cite{snake} the spinons
 become unconfined on  
 the gapless line separating the two phases of the model
 (see Fig. 3.). Since in the two massive phases 
 confinement definitely takes place, we would expect
 that the triplet spectral weight decreases as the critical
 line is approached and 
ultimately goes to zero. In addition, at the point of deconfinement 
 the same energy is required to create triplet and singlet
 excitations. Therefore, close to the critical line, a low-energy
 singlet excitation must appear in the spectrum. Within the
 strong-coupling approach a singlet   can be viewed
 as a collective bound state of two triplets \cite{us2}. 
 While in the spin ladder without dimerization
 the singlet bound state is high in energy (small binding energy), 
  in the present model we expect it to become a truly low-energy
state (strong triplet binding). We leave these issues for a future study
 \cite{us4}.

 While this manuscript was being prepared for publication
 we became aware of two recent preprints devoted to the same model
 \cite{recent}. Both of these works have studied the phase diagram
 numerically via exact diagonalization
 and their results for the location of the critical line are
 in agreement with ours.

\acknowledgements

We would like to thank  O.P. Sushkov for numerous  
stimulating discussions  and critical reading
of the manuscript. This work was supported by a grant from the Australian
Research Council.
The computations have been performed on Silicon Graphics Power
Challenge and Convex machines, and we thank the New South Wales
Centre for Parallel Computing for facilities and assistance
with the calculations.

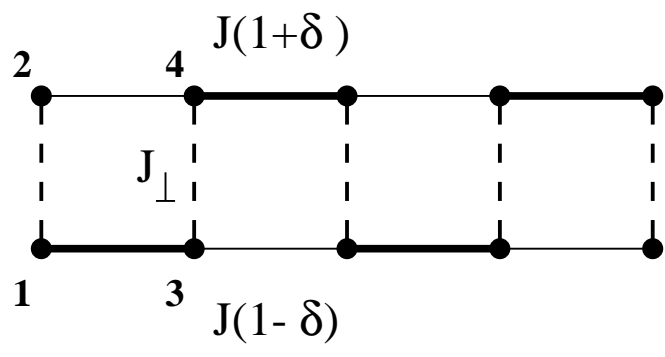
\begin{figure}
\caption{The spin ladder with staggered dimerization.}
\label{fig1}
\end{figure}%

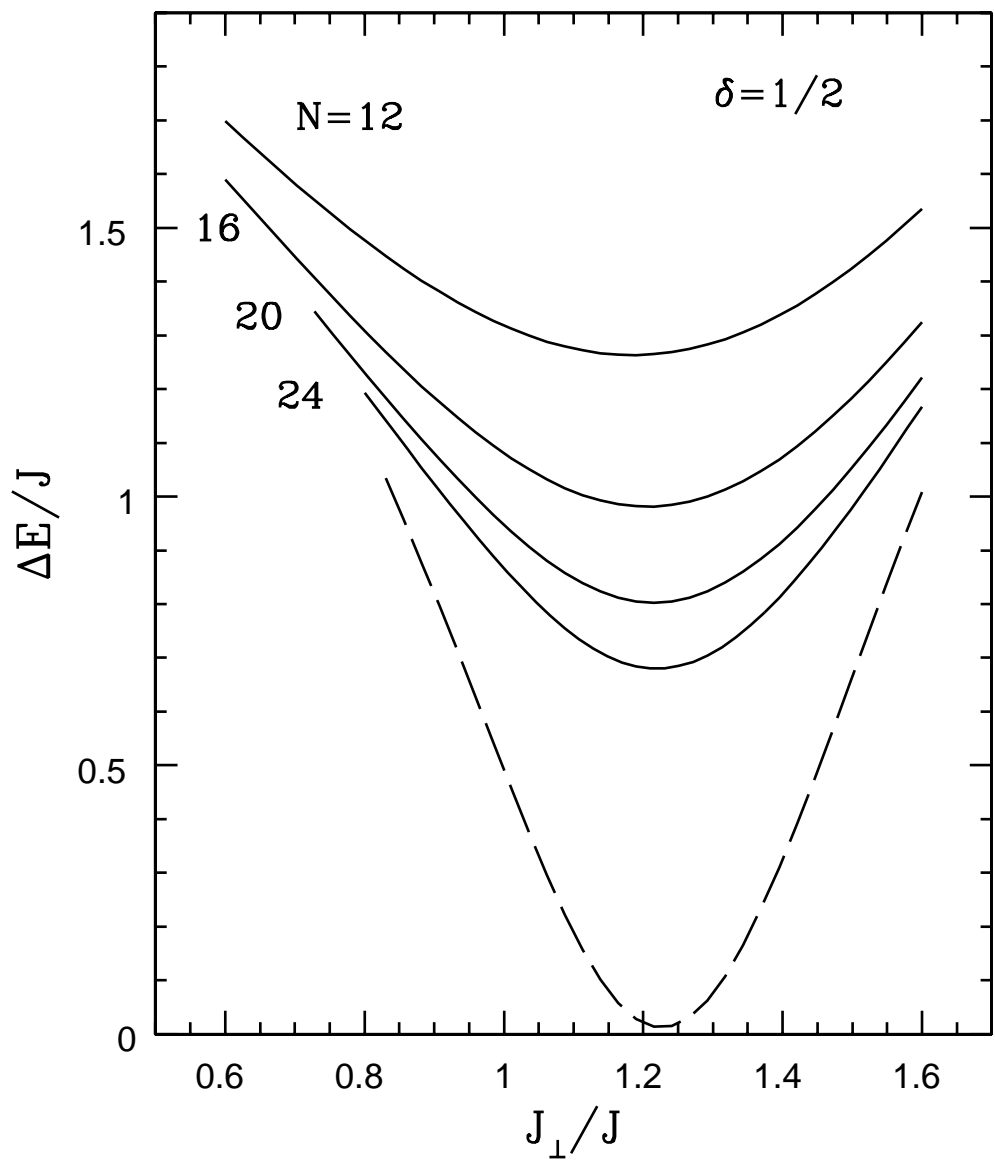
\begin{figure}
\caption{Energy gap for different system sizes (solid lines) and
 its extrapolation to the thermodynamic limit (dashed line).}
\label{ fig2}
\end{figure}%

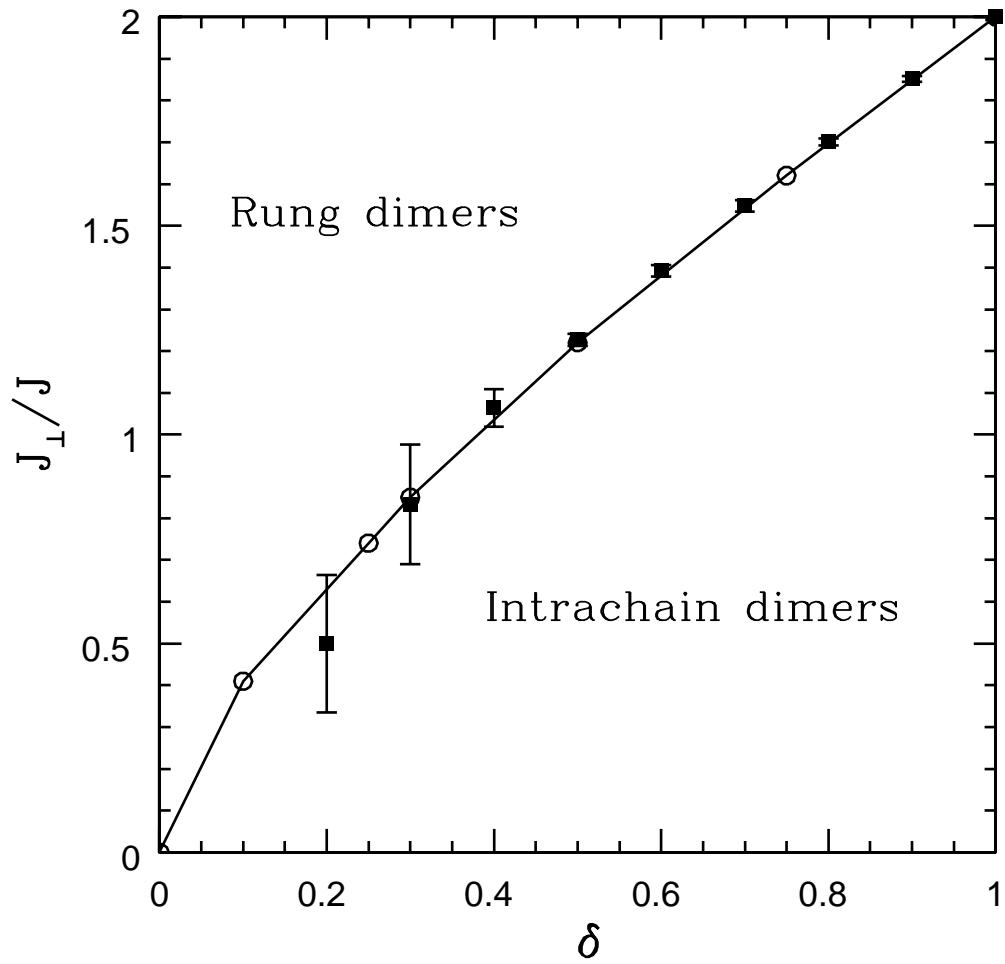
\begin{figure}
\caption{Phase diagram of the model. Open circles are the exact
diagonalization data, where the circle diameter represents the 
error in the determination of the critical point. Solid squares
 with error bars are the critical points obtained by Dlog Pad\'e 
approximation of the dimer series.}
\label{ fig3}
\end{figure}%

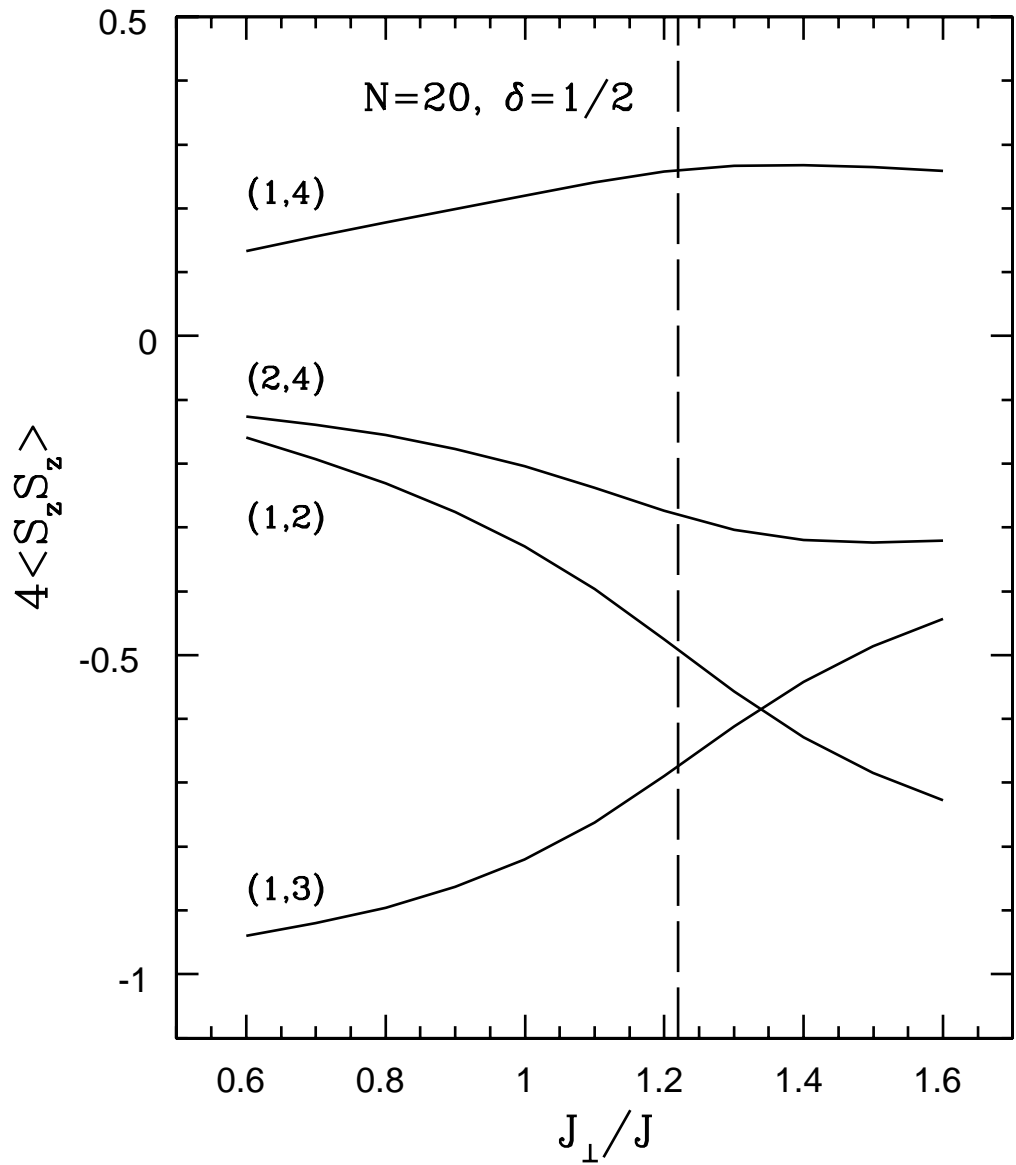
\begin{figure}
\caption{ Spin-spin correlations $\langle S^z_{i} S^z_{j}
\rangle$ (where (i,j) are defined in Fig. 1.) 
obtained by exact diagonalization of a system of $N=20$
 spins. The dashed line represents the location of the critical
 point.}
\label{fig4}
\end{figure}%

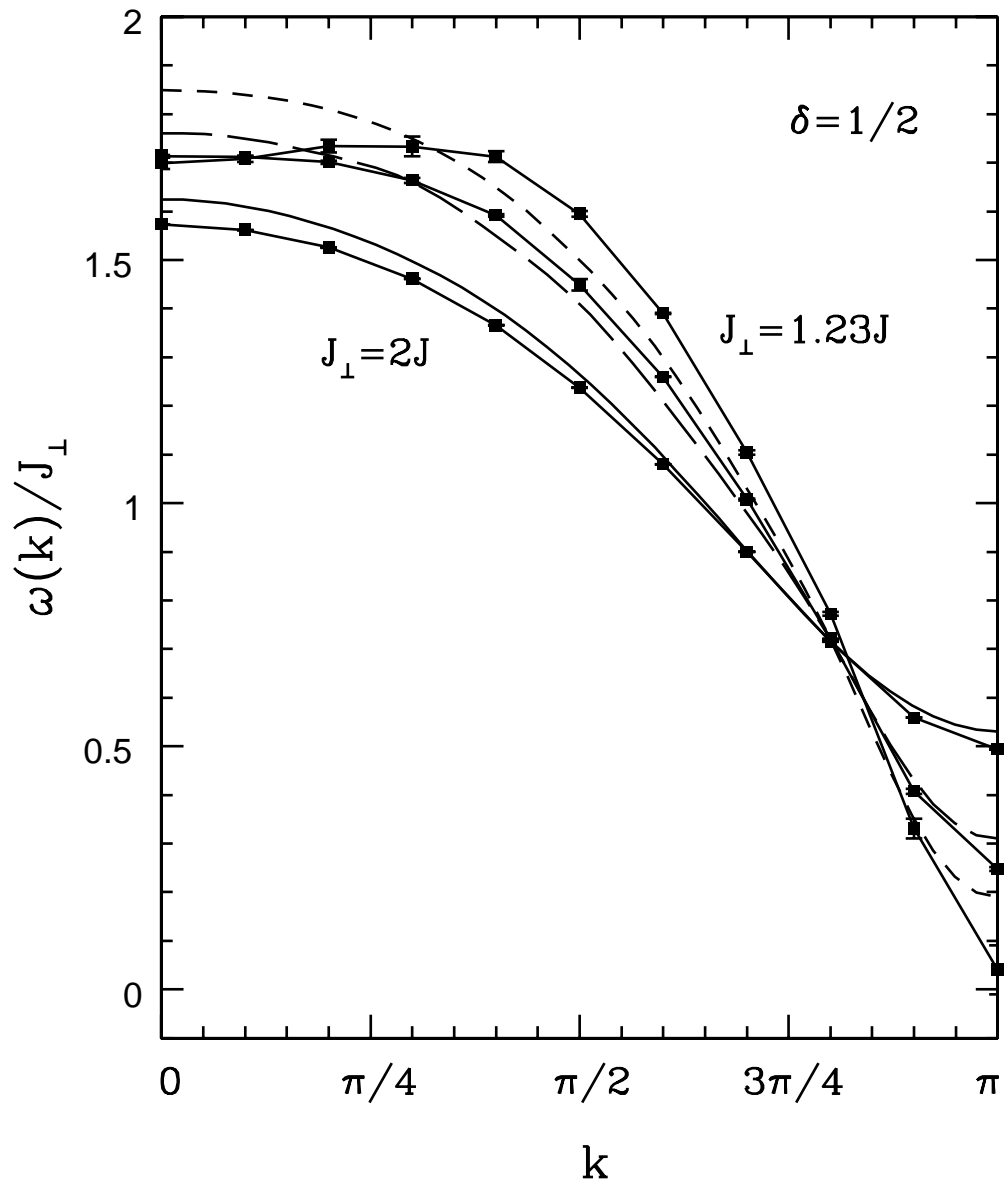
\begin{figure}
\caption{Triplet energy spectrum obtained by the dimer series 
expansion (solid squares connected by solid lines) for
 $J_{\perp}/J = 2, 1.43, 1.23$ (upper, lower and middle curve
 at $k=\pi$, respectively). The corresponding spectra,
 obtained from the poles of Eq.(15) are drawn, respectively, with solid,
long-dashed and short-dashed lines.}
\label{fig5 }
\end{figure}%

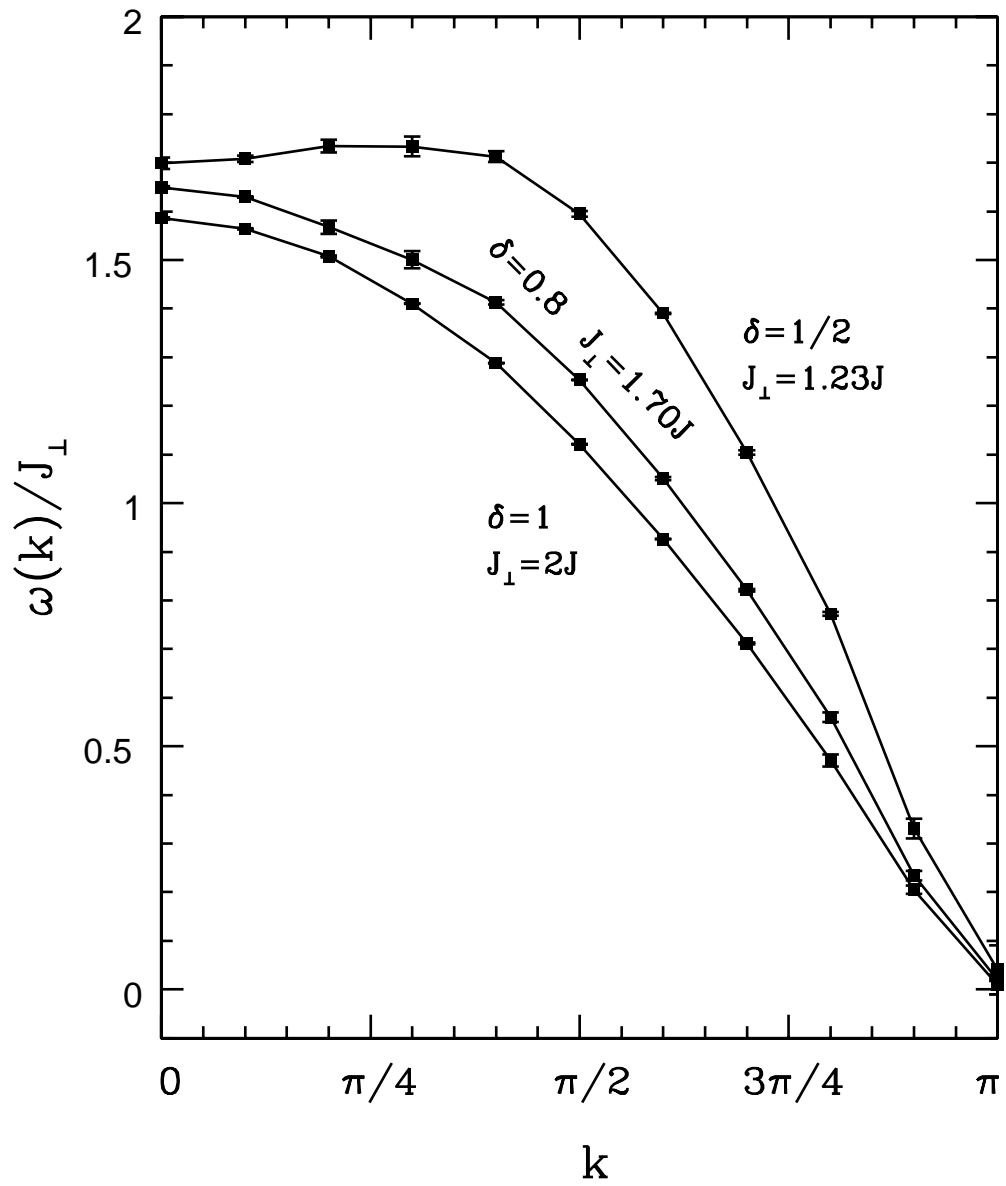
\begin{figure}
\caption{Same as Fig.5. for several values of $(\delta, J_{\perp})$
 on the gapless line.}
\label{fig6 }
\end{figure}%

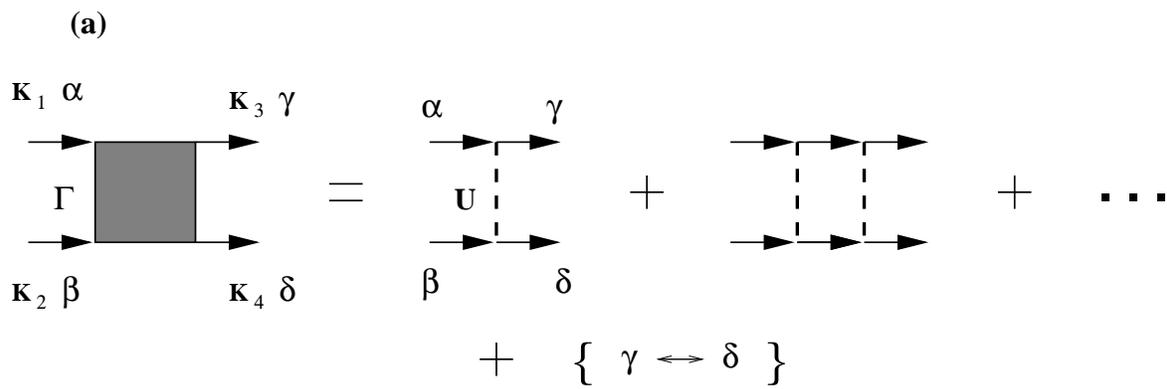
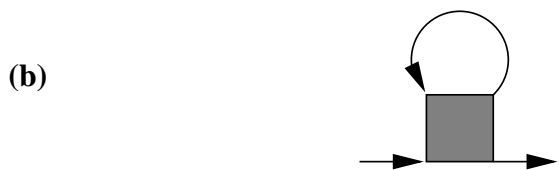
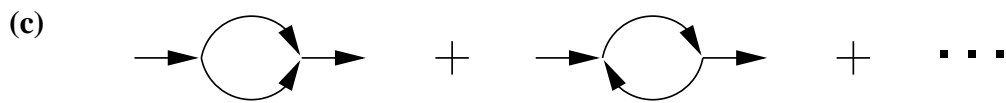
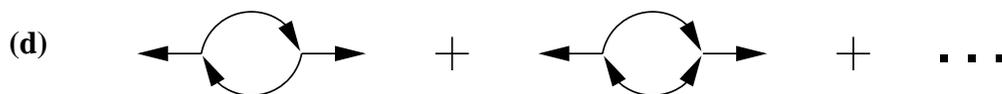
\begin{figure}
\caption{(a) Resummation of the ladder series for the
 scaterring amplitude $\Gamma(k,\omega)$, where
 $k$($\omega$) is the total incoming momentum (energy),
 $k = k_{1} + k_{2}$.  (b) The normal self-energy
 corresponding to $\Gamma$. (c) Lowest order diagrams for
 the normal self-energy   arising
 from the three-particle term $H_{\delta}$. (d) Same as (c) for
 the anomalous self-energy. The dots represent all other possible
 diagrams with different internal loops. }
\label{fig7 }
\end{figure}%


\begin{references}
\bibitem{remark}
The spin operators are related to the triplet  operators
 via the transformation \cite{subir}: $S_{(1,2),i}^{\alpha} = 
1/2(\pm t_{\alpha i}^{\dag} \pm t_{\alpha i} 
+ i\epsilon_{\alpha \beta \gamma} t_{\beta i}^{\dag} t_{\gamma i})$,
 $\alpha = x,y,z$.
\bibitem{remark1} To lowest (one-loop) order, this term leads
 to the renormalization \cite{us2}: $A_{k} \rightarrow
A_{k} + 2J \sum_{q} v_{q}^{2} \cos q$, $B_{k} \rightarrow
 B_{k} - 2J \sum_{q} v_{q}u_{q} \cos q$, which is understood
 in all formulas in the text.
\end{references}
\end{document}